\documentclass[aps,prl,twocolumn,groupedaddress,showpacs]{revtex4}
\usepackage{epsf}

\begin{document}

\title{Coherence Constraint on the Existence of Cosmic Defects}

\author{Jiun-Huei Proty Wu}
\email[]{jhpw@phys.ntu.edu.tw}
\homepage[]{http://jhpw.phys.ntu.edu.tw}
\affiliation{Department of Physics and Institute of Astrophysics,
  National Taiwan University\\
No.1 Sec.4 Roosevelt Road, Taipei 10617, Taiwan}
\date{\today}

\begin{abstract}
We devise a measure $n(r)$ to quantify
a robust feature in the coherent cosmic perturbations,
namely the narrowness of the first peak 
in the power spectrum of the Cosmic Microwave Background (CMB) anisotropy.
A maximum-likelihood analysis
using the WMAP data shows that 
the power fraction of any defect models at the first peak
is persistently less than 4.5\% at the 68\% confidence.
Our approach and results are insensitive to the
uncertainties 
in the theoretical predictions for defect models,
and thus the most robust to date.
We show that
to convert these results into 
realistic constraints on theories such as
inflation, string cosmology, and SUSY GUTs,
a more robust study for the cosmic-string-induced CMB
around the degree scale is required.
\end{abstract}


\maketitle


\noindent
{\bf A.\ Introduction}:
Cosmic defects have been an interesting subject for different reasons
through different decades.
They were first introduced around the 70's as an inevitable consequence of
various theories such as the quantum field theory and the unification theory
\cite{defects:thooft,defects:kibble}.
Then motivated by the anisotropy in the Cosmic Microwave Background (CMB)
observed by COBE in 1992 \cite{COBE:smoot},
they competed with inflation as the dominant mechanism for the formation
of structures in the universe 
\cite{cosmicstrings_cobe:allen,globaldefects,cosmicstrings:albrecht,cosmicstrings:wu_lss,globaldefects:durrer,cosmicstrings:contaldi}.
They started losing this contest in the recent years
when the flood of high-precision data showed great agreement with inflation
(e.g., \cite{maxiboom:jaffe,wmap:para}).
Nevertheless,
defects have been still playing a key role
in testing some of the fundamental physics,
especially those related to inflation and string cosmology.
In general,
for example,
various defects may form as hybrid inflation ends 
\cite{hybridinf:copeland}.
In string theory, colliding branes can not only bring the brane inflation 
to the end but also potentially generate cosmic strings
\cite{cosmicstrings_inf:sarangi,cosmicstrings_inf:dvali}.
In the Supersymmetric (SUSY) Grand Unified Theories (GUTs),
which embed the standard model into the string theory,
cosmic strings can also be formed at the end of hybrid inflation
\cite{cosmicstrings_inf:jeannerot,cosmicstrings_susygut}.
Therefore,
we can use observational data to place limits on the energy density of cosmic
defects, so as to confine the energy scales of the associated physical mechanisms.
In this paper,
we will employ the CMB data for this purpose.

\begin{figure}
  \centering 
  \leavevmode\epsfxsize=8cm \epsfbox{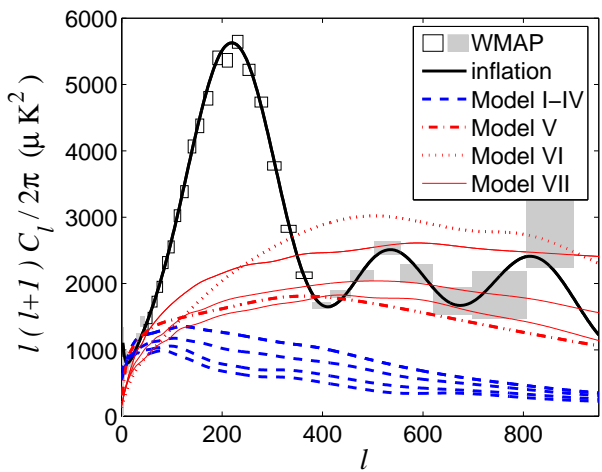}\\
  \caption[]
	  {CMB power spectra of global defects (blue dashed; Models I--IV downwards),
	    local cosmic strings (red dot-dashed, red dotted, and red thin solid),
	    the concordance model (black thick),
	    and the WMAP (boxes; at 1-$\sigma$).
	    The 19 empty boxes are used
	    for the analysis of $n(r)$ and $f^x_{220}$.
	    Model V is normalized with $G\mu=2\times 10^{-6}$;
	    Models VI and VII
	    are both with $G\mu=7\times 10^{-7}$ for the ease of comparison.
}
\label{fig-llcl}
\end{figure}

There are two main types of defects:
global defects (Models I--IV in Tab.~\ref{tab-model})
and local cosmic strings (Models V--VII).
Their resulting CMB power spectra are shown in Fig.~\ref{fig-llcl}.
The CMB power spectrum is defined as $C_\ell=\langle |a_{\ell m}|^2 \rangle$,
where $a_{\ell m}$'s are the multipole-expansion coefficients of
the CMB temperature anisotropy,
and $\ell$ is the multipole number.
By comparing these results side by side,
we find that
the results for the $O(4)$ textures (Model III) in Ref.\cite{globaldefects} 
and \cite{globaldefects:durrer} agree well at the 10\% level for $\ell<1000$.
On the other hand, 
the uncertainty in the theoretical predictions for cosmic strings remains large.
Models V\cite{cosmicstrings:pogosian} and VI\cite{cosmicstrings:albrecht}
are both based on a toy model using filaments to mimic strings,
while Model VII\cite{cosmicstrings:contaldi} uses string simulations 
extrapolated from a non-expanding universe.
These predictions disagree not only in shape but also in amplitude.
For example, although based on the same approach,
the predicted $C_\ell$ of Model VI~\cite{cosmicstrings:albrecht} is higher than 
that of Model V~\cite{cosmicstrings:pogosian}
by a factor of 4 and 16 at $\ell=10$ and $700$ respectively 
(for the same string linear energy density $\mu$, where $C_\ell\propto \mu^2$;
see Fig.~\ref{fig-llcl}).
Depending on the velocity of the decay products of the long strings,
the amplitude and the shape of $C_\ell$ may vary significantly
\cite{cosmicstrings:contaldi} 
(see Model VII in Fig.~\ref{fig-llcl}).
These theoretical uncertainties have been overlooked in most of the recent studies 
about the observational constraint 
on $\mu$ and other related model parameters
(e.g., \cite{cosmicstrings:pogosian,cosmicstrings_wmap_sdss:pogosian,cosmicstrings_susygut}),
dramatically weakening the conclusions therein
(e.g., $G\mu\leq 2\times 10^{-7}$ at 99\% confidence 
in Ref.\cite{cosmicstrings:pogosian,cosmicstrings_wmap_sdss:pogosian},
which used WMAP and SDSS data).
Note that work employing full string simulations in an expanding universe
has also been pursued
\cite{cosmicstrings:bouchet_nature,cosmicstrings_cobe:allen,cosmicstrings:wu_lss,cosmicstrings:wu,phdthesis,cosmicstrings:cmb,cosmicstrings_amiba} 
but only for $\ell$ up to a few tens due to the numerical limitation
\cite{cosmicstrings:cmb}.
This $\ell$ range is not suitable for us to investigate 
the observational first peak at $\ell\approx 220$ (see later),
so we will not consider this model here.
It is still worthy mentioning that 
the $C_{10}$ in Ref.\cite{cosmicstrings:cmb}
is about 8 times higher than that in Ref.\cite{cosmicstrings:pogosian}
for the same $\mu$.

In face of the above uncertainties in the theoretical work,
we devise a new strategy to constrain the energy level 
and thus the existence of defects.
This strategy
focuses on a robust feature of the coherent cosmic 
perturbations---the narrowness of the first peak in ${C}_\ell$.

\begin{figure}
  \centering 
  \leavevmode\epsfxsize=8cm \epsfbox{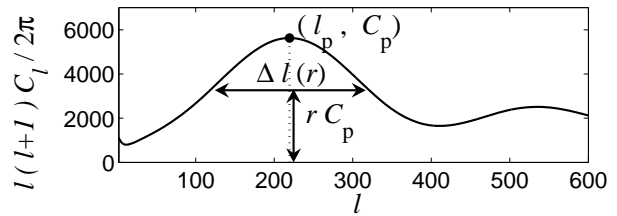}\\
  \caption[]
	  {Illustration for the definition of $n(r)$ in Eq.~(\ref{n}).
	  }
	  \label{fig-def}
\end{figure}

\noindent
{\bf B.\ Quantifying the coherence}:
We first define ${\cal C}_{\ell}= \ell(\ell+1) C_\ell/2\pi$.
If the apex of the first peak in ${\cal C}_{\ell}$ is located at $\ell_{\rm p}$
with the height ${\cal C}_{\rm p}\equiv {\cal C}(\ell_{\rm p})$,
and
the width of the peak at the height $r{\cal C}_{\rm p}$ ($0< r \leq 1$) is $\Delta\ell(r)$
(see Fig.~\ref{fig-def}),
then the new measure is defined as
\begin{equation}
  \label{n}
  n(r)=\frac{\Delta\ell(r)}{\ell_{\rm p}}.
\end{equation}
As will be justified and explained later,
the shape of the function $n(r)$ is determined by the nature of 
whether the CMB perturbations are coherent or not.
It depends only weakly on the geometry of the universe,
because a change in geometry will stretch 
both  $\Delta\ell$ and  $\ell_{\rm p}$ in the same way,
so that their ratio $n(r)$ remains very much unchanged.

\begin{table} 
  \caption{\label{tab-model}Cosmic defect models investigated here,
    and their power fraction $f^x_{220}$ at $\ell=220$, 
    their temperature anisotropy $\sqrt{{\cal C}^x_\ell}$ at $\ell=220$ and $10$
    (in $\mu{\rm K}$),
    as constrained by WMAP, all at the $68\%$ confidence.}
  \begin{tabular}{|| r | l | c | c | c | c ||}
    \hline
    \hline
    $x$ \quad & \quad Models & ~$f^x_{220}$
    & $\sqrt{{\cal C}^x_{220}}$ & $\sqrt{{\cal C}^x_{10}}$ & constraints \\
    \hline
    \hline
    I &  strings\cite{globaldefects} & $\leq 4.5\%$ 
    & $\leq 16$ & $\leq 13$ & - \\
    II &  monopoles\cite{globaldefects} &  $\leq 4.4\%$ 
    & $\leq 16$ & $\leq 14$ & - \\
    III & textures\cite{globaldefects,globaldefects:durrer} &  $\leq 3.7\%$ 
    & $\leq 14$ & $\leq 15$ &  $\epsilon \leq 8.2\times 10^{-6}$ \\
    IV & $O(6)$  text.\cite{globaldefects} &  $\leq 3.0\%$
    & $\leq 13$ & $\leq 14$ & - \\
    \hline
    V &  loc.~cos.~str.\cite{cosmicstrings:pogosian} &  $\leq 3.4\%$ 
    & $\leq 14$ & $\leq 10$ &  $G\mu\leq 7\times 10^{-7}$ \\ 
    VI &  loc.~cos.~str.\cite{cosmicstrings:albrecht} &  $\leq 2.1\%$ 
    & $\leq 11$ & $\leq 4.6$ &  $G\mu\leq 2\times 10^{-7}$\\ 
    VII &  loc.~cos.~str.\cite{cosmicstrings:contaldi} & -
    & - & - & -\\
    \hline
    \hline
  \end{tabular}
\end{table}

\begin{figure}
  \centering 
  \leavevmode\epsfxsize=8cm \epsfbox{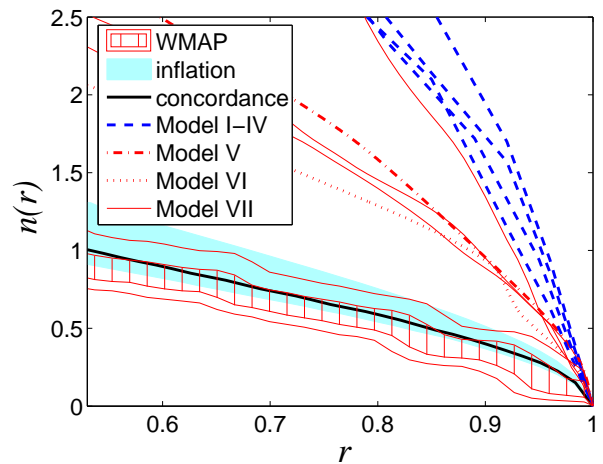}\\
  \caption[]
	  {The $n(r)$ of the concordance model and the defects
	    (same line styles as in Fig.~\ref{fig-llcl}).
	    Additionally, the inflationary models spam over 
	    the blue shaded region,
	    and the red lined regions are
	    the 68\% and 95\% C.R.'s from WMAP.
	  }
	  \label{fig-nr}
\end{figure}

To justify the robustness of this coherence feature in $n(r)$,
we consider the inflationary models
for a reasonably large region in the parameter space:
$0\leq\Omega_\Lambda\leq 1$ (energy density parameter for cosmological constant),
$0.1\leq\Omega_{\rm c}\leq 1$ (for cold dark matter), 
$0.01\leq\Omega_{\rm b}\leq 0.2$ (for baryons), 
$0.5\leq h\leq 0.9$ 
($H_0=100h~{\rm km~s^{-1}Mpc^{-1}}$ where $H_0$ is the Hubble parameter today), 
$0.8\leq n_{\rm s}\leq 1.2$ (scalar spectral index),
$0\leq\tau\leq 0.25$ (optical depth).
Note that this gives 
$0.11\leq \Omega_{\rm t}\leq 2.2$ (total energy density parameter).
First we compute all the CMB power spectra \cite{cmbfast}
within this parameter range,
and then the resulting $n(r)$ for each spectrum. 
These $n(r)$'s occupy the blue shaded region in Fig.~\ref{fig-nr}.
It is clear that
the coherence nature of the inflationary models confines the $n(r)$
within a very limited region on the $r$-$n$ plane.

On the other hand, we compute the $n(r)$ for the defect models 
in Tab.~\ref{tab-model}.
Obviously the peaks in ${\cal C}_\ell$ for these incoherent models
are much broader (Fig.~\ref{fig-llcl}),
resulting in an $n(r)$ of at least twice higher at a given $r$ (Fig.~\ref{fig-nr}).
A comparison between these lines and the shaded blue region shows
the capability of $n(r)$ in discriminating the coherent and
the incoherent models.
This discrimination is independent of the cosmological parameters.


Theoretically
this coherence feature in the $n(r)$ can be manifested
by the following intuitive approach. 
In the synchronous gauge for a perturbation theory \cite{defects_incoh,phdthesis},
it is straight forward to show that with inflation
the power spectrum of photon energy contrasts at the last-scattering epoch
$\eta_{\rm ls}$ can be approximated by the form:
\begin{equation}
  P(k) \propto
  k^{n_{\rm s}} 
  \left[
    \frac{\cos\left(k\eta_{\rm ls}/\sqrt{3}\right)}
	 {1+\left(k\eta_{\rm ls}\right)^2}
	 e^{-(k\eta_{_D})^2}
	 \right]^2,
  \label{Pk}
\end{equation}
where $k$ is the comoving wave number.
The exponential term accounts for the photon-diffusion damping,
with $\eta_{_D}$ typically one order below $\eta_{\rm ls}$ \cite{cosperturb:hu}.
The cosine oscillations are due to the coherence nature of the perturbations,
resulting from the fact that 
the inflationary perturbations on the same scale ($k^{-1}$)
entered the horizon at the same time ($\sim k^{-1}$) and thus oscillated
coherently.
On the other hand,
for models with only defects,
the perturbations on the same scale were seeded over a range of time
after the horizon crossing,
and thus oscillated incoherently. 
These incoherent oscillations effectively remove the cosine dependence 
in Eq.~(\ref{Pk}), 
leaving only a broad peak around the scale $\eta_{\rm ls}$.
We then compute
$C_\ell \propto \int P(k) j_\ell(k\eta_0) k^2dk$,
where $\eta_0$ is the comoving radius of the last-scattering sphere and
$j_\ell$ is the spherical Bessel function.
We find that
the resulting $n(r)$ for the coherent case is slightly
lower than the blue shaded region in Fig.~\ref{fig-nr},
due to the omission of the Doppler effect 
and the Integrated Sachs-Wolfe effect.
These two effects are sub-dominant in the first peak and thus in $n(r)$.
It is important to note that 
the cosine dependence in $P(k)$ gives not only the periodic peaks
in $C_\ell$ as commonly known,
but also the narrowness of the first peak
as we try to advocate here.
On the other hand,
the incoherent case 
leads to a much broader peak in $C_\ell$, 
resulting in an $n(r)$ of at least twice larger than the coherent case.

The cosmic perturbations may be attributed to
both coherent and incoherent mechanisms.
Given the fact that the former is the dominant,
an increasing fraction of the latter in a mix model
will broaden the width of the first peak,
resulting in a higher value of $n(r)$. 
In the following,
we will invoke this feature
to constrain the fraction of the incoherent perturbations
and thus the defects.

\noindent
{\bf C.\ Observational constraint}:
From observations, we can estimate $n(r)$.
Among the large amount of CMB results 
that measure the first peak in ${\cal C}_\ell$,
we choose the WMAP \cite{wmap:pow}
because it is the most stringent to date.
We will denote its result as
$(\ell_{i(\rm l)}, \ell_{i(\rm r)}, {\cal C}_i, \sigma_i)$,
where $i$ indicates the $i$-th $\ell$ bin,
$\ell_{i(\rm l)}$ and $\ell_{i(\rm r)}$ denote respectively the smallest
and the largest $\ell$ within the bin,
${\cal C}_i$ is the binned power, 
and $\sigma_i$ is the 1-$\sigma$ error in ${\cal C}_i$.
For a given point $A(\ell'_{\rm p}, {\cal C}'_{\rm p})$ 
on the $\ell$-${\cal C}_\ell$ plane,
the probability that it is the apex of the first peak is
\begin{equation}
  \label{eq-P1}
  P_1(A) =
  pdf({\cal C}'_{\rm p}; {\cal C}_a, \sigma_a)
  \prod_{i\neq a} cdf({\cal C}'_{\rm p}; {\cal C}_i, \sigma_i),
\end{equation}
where $a$ denotes the $a$-th $\ell$ bin within which $\ell'_{\rm p}$ lies
(i.e.\ $\ell'_{\rm p}\in [\ell_{a(\rm l)}, \ell_{a(\rm r)}]$),
$pdf(x; \mu,\sigma)$ is a normal probability distribution function
with mean $\mu$ and standard deviation $\sigma$,
$cdf(x; \mu,\sigma)$ is a normal cumulative probability distribution function.
Then at the height $r{\cal C}'_{\rm p}$ ($0< r\leq 1$),
the probablility for point $B(\ell'_{\rm l}, r{\cal C}'_{\rm p})$ 
being the left-end point of the peak and 
$\Delta \ell'$ being the width of the peak is
\begin{eqnarray}
  \label{eq-P2}
  && P_2(B,\Delta\ell'|A)  = 
  \prod_{\ell_{i(\rm r)}<\ell'_{\rm l} \text{ or } \ell'_{\rm l}+\Delta\ell'<\ell_{i(\rm l)}}
  cdf(r{\cal C}'_{\rm p}; {\cal C}_i, \sigma_i)   \nonumber\\
  &&
  \times \prod_{\ell'_{\rm l} < \ell_{i(\rm l)} \text{ \& } \ell_{i(\rm r)}<\ell'_{\rm l}+\Delta\ell'} 
  \left[1-cdf(r{\cal C}'_{\rm p}; {\cal C}_i, \sigma_i)\right]    \nonumber\\
  &&
  \times I(\ell'_{\rm l}; r{\cal C}'_{\rm p}, \ell_{b{\rm (l)}}, \ell_{b{\rm (r)}})
  I(\ell'_{\rm l}+\Delta\ell'; r{\cal C}'_{\rm p}, \ell_{c{\rm (r)}}, \ell_{c{\rm (l)}}),
  \quad
\end{eqnarray}
where $b$ and $c$ denote respectively the $\ell$ bins 
within which the $\ell'_{\rm l}$ and the $\ell'_{\rm l}+\Delta\ell'$ are, and
$I(x; r{\cal C}'_{\rm p}, \ell_{i{\rm (1)}}, \ell_{i{\rm (2)}})$
is a linear-interpolation function
between $(\ell_{i{\rm (1)}}, 1-cdf(r{\cal C}'_{\rm p};{\cal C}_i,\sigma_i))$ and 
$(\ell_{i{\rm (2)}}, cdf(r{\cal C}'_{\rm p};{\cal C}_i,\sigma_i))$ according to $x$.
Here the condition
$\ell'_{\rm l}<\ell'_{\rm p}<\ell'_{\rm l}+\Delta\ell'$
is also imposed.
Finally, given the observational data,
the probability for a point on the $r$-$n$ plane to be present is
\begin{equation}
  \label{eq-P}
  P(r,n) = \sum_{A, \Delta\ell'/\ell'_{\rm p}=n} P_1(A) P_2(B,\Delta\ell'|A).
\end{equation}
For each $r$, we compute the 68\% and the 95\% confidence regions (C.R.) for $n$, 
as shown in Fig.~\ref{fig-nr}.
Note that for the WMAP data,
we use only the 19 bins ($i=12$--$30$; $45\leq\ell\leq 379$) 
that best confine the first peak. 
These bins are shown as the empty boxes in Fig.~\ref{fig-llcl}.

It is quite clear in Fig.~\ref{fig-nr} that
the observational data strongly favor the inflationary models,
regardless of the geometry of the universe.
Thus a defect-dominated universe with a very closed geometry,
which serves to shift the first peak from several hundreds to around 220,
is totally inconsistent with the observations.
Also plotted in Fig.~\ref{fig-nr} is the $n(r)$ of
the concordance model based on
the data of
WMAP, CBI, ACBAR, 2dF, Ly-alpha \cite{wmap:para}.
Its power spectrum is shown 
in Fig.~\ref{fig-llcl}.
Its $n(r)$ is roughly within the 68\% C.R.\ of the data.
We also note from Fig.~\ref{fig-nr} 
that on average the data favor a slightly narrower peak 
than what is predicted by the inflationary models.

To see how stringent the observed narrowness of the first peak can constrain the
existence of cosmic defects,
we further consider the following mixed power spectrum:
\begin{equation}
  \label{f220}
	{\cal C}^{\rm tot}_\ell =
	A \left[
	f^x_{220} \frac{{\cal C}^{\rm inf}_{220}}{{\cal C}^x_{220}}{\cal C}^x_\ell
	+
	(1 - f^x_{220}) {\cal C}^{\rm inf}_\ell
	\right],
	x=\text{I--VI},
\end{equation}
where the superscripts `inf' and `$x$' denote the contribution 
from inflation
and
from the defect models in Tab.~\ref{tab-model}
respectively.
Model VII is not considered here 
because Ref.\cite{cosmicstrings:contaldi} does not provide information
about the dependence of the ${\cal C}_\ell$ on the cosmological parameters.
Here $f^x_{220}$ is nothing but the power fraction of the defects
at the first peak ($\ell_{\rm p}\approx 220$), 
and $A$ is an arbitrary normalization factor.
We use the WMAP data to constrain $f^x_{220}$,
again using only the 19 $\ell$ bins. 
A standard maximum-likelihood analysis is performed
for the parameters of 
$f^x_{220}$, $A$, $\Omega_\Lambda$, $\Omega_{\rm c}$,
$\Omega_{\rm b}$, $h$, $n_{\rm s}$, $\tau$,
with the flat-geometry constraint that
$\Omega_{\rm t}=\Omega_\Lambda+\Omega_{\rm c}+\Omega_{\rm b}=1$.
We then marginalize over all the parameters except for $f^x_{220}$,
leaving only a one-dimensional likelihood ${\cal L}(f^x_{220})$.
At the 68\% confidence level (C.L.),
the estimated values of $f^x_{220}$ are shown in Tab.~\ref{tab-model}.
Two robust features in all results for different models are that
the likelihood  ${\cal L}(f^x_{220})$ peaks below $f^x_{220}=1.2\%$,
and that $f^x_{220}\leq 4.5\%$
(or equivalently $\sqrt{{\cal C}^x_{220}} \leq 16\mu {\rm K}$)
at the 68\% C.L.
This means that
the defect models contribute at most 4.5\%,
preferably below 1.2\%,
in the observed ${\cal C}_{220}$,
or equivalently that
their contribution to the CMB anisotropy around the degree scale
is no more than $16\mu {\rm K}$,
preferably below $8.2\mu {\rm K}$.
The $\sqrt{{\cal C}^x_{10}}$ in Tab.~\ref{tab-model}
is extrapolated from the value of $\sqrt{{\cal C}^x_{220}}$
taking the corresponding model shape of ${\cal C}^x_\ell$.
We see that  $\sqrt{{\cal C}^x_{10}} \leq 15\mu {\rm K}$  (around the COBE scale)
for all models.

\noindent
{\bf D.\ Discussion and Conclusion}:
We emphasize that
although the shape and the amplitude of ${\cal C}^x_\ell$ 
vary a lot among different defect models,
the observational constraint on their $f^x_{220}$ remains persistently below 4.5\%.
When we replace the ${\cal C}^x_\ell$ in Eq.~(\ref{f220})
with a ${\cal C}_\ell$ which has $d{\cal C}_\ell/d\ell =m$,
the same conclusion persists for $-10\leq m\leq 10$.
This indicates that
any further improvement on the detailed predictions of ${\cal C}^x_\ell$
for defect models can only marginally affect our results here.
In theory this is naturally due to the fact that
the predicted ${\cal C}^x_\ell$'s are lack of curvature 
around the observational first peak at $\ell \approx 220$,
so that a larger $f^x_{220}$ will result in a higher $n(r)$,
which potentially violates the observational data.

Our results can be converted to place limits on the parameters in various theories.
For example, 
the result $\sqrt{{\cal C}^{\rm III}_{10}}\leq 15\mu {\rm K}$ 
can be converted to
$\epsilon=4\pi G\phi_0^2 \leq 8\times 10^{-6}$,
leading to a constraint on the energy level of symmetry breaking
$\phi_0\leq 9.9\times 10^{15}$GeV.
This is consistent with, though more conservative than,
the result in Ref.\cite{textures:wmap},
where
the full WMAP data were used to yield
$f^{\rm III}_{10}\leq 13\%$,
or equivalently $\sqrt{{\cal C}^{\rm III}_{10}}\leq 10\mu {\rm K}$, at 95\% C.L.

For local cosmic strings,
naively we can take the more conservative result for $\mu$, 
from Model V (see Tab.~\ref{tab-model}),
to yield a constraint on the symmetry-breaking scale
$\eta \sim \mu^{1/2}\leq 10^{16}$GeV.
In the context of string theory,
where the strings are D-branes,
it places an upper limit onto the superstring scale
$M_{\rm s}\approx (\mu/2)^{1/2}\leq 7\times 10^{15}$GeV.
In the context of SUSY GUTs,
a combination of our further deduced result 
$Q^{V}_{rms-PS}\equiv (\delta T)^V_Q\leq 4.4\mu {\rm K}$ 
and the study in Ref.\cite{cosmicstrings_susygut}
yields an upper limit on the mass scale of the F-term inflation
$M\leq 2\times 10^{15}$GeV, 
and on the superpotential coupling
$\kappa\leq 2\times 10^{-5}$.
For the D-term inflation,
we get constraint on the gauge coupling $g\leq 1.4\times 10^{-2}$,
and on the superpotential coupling $\lambda \leq 2.2\times 10^{-5}$.
We emphasize that
all these constraints still carry considerable uncertainties
due to the large uncertainties 
in the predicted ${\cal C}_{220}$ for strings,
as addressed in the introduction.
This uncertainty is propagated into 
the constraint on $\mu$ and thus on the deduced parameters.
This is readily seen by the inconsistency between
$G\mu\leq 7\times10^{-7}$ and $2\times 10^{-7}$ 
for Models V and VI respectively (see Tab.~\ref{tab-model}).
Therefore 
a more robust predictions for strings around the degree scale
will dramatically settle this uncertainty.
For the same reason,
whether or not 
the values of $G\mu=5.9\times 10^{-7}$ \cite{cosmicstrings_lens:sazhin} 
and $4\times 10^{-7}$ \cite{cosmicstrings_lens:schild}
that were required to explain the recent anomalous observations
in the gravitational lens systems
is consistent with the current CMB constraint
remains an open question.
Nevertheless, we are confident that
$\sqrt{{\cal C}_{220}^{\rm str}}\leq 14\mu {\rm K}$.

In conclusion,
we invoke a coherence 
feature---the narrowness of the first peak in ${\cal C}_\ell$---to
constrain the existence of defects.
We show that 
the power fraction of any defect models at the first peak is 
persistently less than 4.5\% at the 68\% C.L.,
in spite of the large uncertainties in the current theoretical study
(see Tab.~\ref{tab-model} for the more detailed results).
Hence our results should be 
the most objective and thus the most robust to date.



\begin{acknowledgments}
We acknowledge the support from the National Science Council 
of Taiwan (NSC 92-2112-M-002-047; NSC 93-2112-M-002-015).
\end{acknowledgments}


\begin{thebibliography}{30}
\expandafter\ifx\csname natexlab\endcsname\relax\def\natexlab#1{#1}\fi
\expandafter\ifx\csname bibnamefont\endcsname\relax
  \def\bibnamefont#1{#1}\fi
\expandafter\ifx\csname bibfnamefont\endcsname\relax
  \def\bibfnamefont#1{#1}\fi
\expandafter\ifx\csname citenamefont\endcsname\relax
  \def\citenamefont#1{#1}\fi
\expandafter\ifx\csname url\endcsname\relax
  \def\url#1{\texttt{#1}}\fi
\expandafter\ifx\csname urlprefix\endcsname\relax\def\urlprefix{URL }\fi
\providecommand{\bibinfo}[2]{#2}
\providecommand{\eprint}[2][]{\url{#2}}

\bibitem[{\citenamefont{t'Hooft}(1974)}]{defects:thooft}
\bibinfo{author}{\bibfnamefont{G.}~\bibnamefont{t'Hooft}},
  \bibinfo{journal}{Nucl.\ Phys.\ B} \textbf{\bibinfo{volume}{79}},
  \bibinfo{pages}{276} (\bibinfo{year}{1974}).

\bibitem[{\citenamefont{Kibble}(1976)}]{defects:kibble}
\bibinfo{author}{\bibfnamefont{T.~W.~B.} \bibnamefont{Kibble}},
  \bibinfo{journal}{J.\ Phys.\ A} \textbf{\bibinfo{volume}{9}},
  \bibinfo{pages}{1387} (\bibinfo{year}{1976}).

\bibitem[{\citenamefont{{Smoot {\em et al.}}}(1992)}]{COBE:smoot}
\bibinfo{author}{\bibfnamefont{G.~F.} \bibnamefont{{Smoot {\em et al.}}}},
  \bibinfo{journal}{Astroph.\ J.\ Lett.} \textbf{\bibinfo{volume}{396}}
  (\bibinfo{year}{1992}).

\bibitem[{\citenamefont{Allen et~al.}(1996)\citenamefont{Allen, Caldwell,
  Shellard, Stebbins, and Veeraraghavan}}]{cosmicstrings_cobe:allen}
\bibinfo{author}{\bibfnamefont{B.}~\bibnamefont{Allen}},
  \bibinfo{author}{\bibfnamefont{R.~R.} \bibnamefont{Caldwell}},
  \bibinfo{author}{\bibfnamefont{E.~P.~S.} \bibnamefont{Shellard}},
  \bibinfo{author}{\bibfnamefont{A.}~\bibnamefont{Stebbins}}, \bibnamefont{and}
  \bibinfo{author}{\bibfnamefont{S.}~\bibnamefont{Veeraraghavan}},
  \bibinfo{journal}{Phys.\ Rev.\ Lett.} \textbf{\bibinfo{volume}{77}},
  \bibinfo{pages}{3061} (\bibinfo{year}{1996}).

\bibitem[{\citenamefont{Pen et~al.}(1997)\citenamefont{Pen, Seljak, and
  Turok}}]{globaldefects}
\bibinfo{author}{\bibfnamefont{U.~L.} \bibnamefont{Pen}},
  \bibinfo{author}{\bibfnamefont{U.}~\bibnamefont{Seljak}}, \bibnamefont{and}
  \bibinfo{author}{\bibfnamefont{N.}~\bibnamefont{Turok}},
  \bibinfo{journal}{Phys.\ Rev.\ Lett.} \textbf{\bibinfo{volume}{79}},
  \bibinfo{pages}{1611} (\bibinfo{year}{1997}).

\bibitem[{\citenamefont{Albrecht et~al.}(1997)\citenamefont{Albrecht, Battye,
  and Robinson}}]{cosmicstrings:albrecht}
\bibinfo{author}{\bibfnamefont{A.}~\bibnamefont{Albrecht}},
  \bibinfo{author}{\bibfnamefont{R.~A.} \bibnamefont{Battye}},
  \bibnamefont{and} \bibinfo{author}{\bibfnamefont{J.}~\bibnamefont{Robinson}},
  \bibinfo{journal}{Phys.\ Rev.\ Lett.} \textbf{\bibinfo{volume}{79}},
  \bibinfo{pages}{4736} (\bibinfo{year}{1997}).

\bibitem[{\citenamefont{Avelino et~al.}(1998)\citenamefont{Avelino, Shellard,
  Wu, and Allen}}]{cosmicstrings:wu_lss}
\bibinfo{author}{\bibfnamefont{P.~P.} \bibnamefont{Avelino}},
  \bibinfo{author}{\bibfnamefont{E.~P.~S.} \bibnamefont{Shellard}},
  \bibinfo{author}{\bibfnamefont{J.~H.~P.} \bibnamefont{Wu}}, \bibnamefont{and}
  \bibinfo{author}{\bibfnamefont{B.}~\bibnamefont{Allen}},
  \bibinfo{journal}{Phys.\ Rev.\ Lett.} \textbf{\bibinfo{volume}{81}},
  \bibinfo{pages}{2008} (\bibinfo{year}{1998}).

\bibitem[{\citenamefont{Durrer et~al.}(1999)\citenamefont{Durrer, Kunz, and
  Melchiorri}}]{globaldefects:durrer}
\bibinfo{author}{\bibfnamefont{R.}~\bibnamefont{Durrer}},
  \bibinfo{author}{\bibfnamefont{M.}~\bibnamefont{Kunz}}, \bibnamefont{and}
  \bibinfo{author}{\bibfnamefont{A.}~\bibnamefont{Melchiorri}},
  \bibinfo{journal}{Phys.\ Rev.\ D} \textbf{\bibinfo{volume}{59}},
  \bibinfo{pages}{123005} (\bibinfo{year}{1999}).

\bibitem[{\citenamefont{Contaldi et~al.}(1999)\citenamefont{Contaldi,
  Hindmarsh, and Magueijo}}]{cosmicstrings:contaldi}
\bibinfo{author}{\bibfnamefont{C.}~\bibnamefont{Contaldi}},
  \bibinfo{author}{\bibfnamefont{M.}~\bibnamefont{Hindmarsh}},
  \bibnamefont{and} \bibinfo{author}{\bibfnamefont{J.}~\bibnamefont{Magueijo}},
  \bibinfo{journal}{Phys.\ Rev.\ Lett.} \textbf{\bibinfo{volume}{82}},
  \bibinfo{pages}{679} (\bibinfo{year}{1999}).

\bibitem[{\citenamefont{{Jaffe {\em et al.}}}(2001)}]{maxiboom:jaffe}
\bibinfo{author}{\bibfnamefont{A.~H.} \bibnamefont{{Jaffe {\em et al.}}}},
  \bibinfo{journal}{Phys.\ Rev.\ Lett.} \textbf{\bibinfo{volume}{86}},
  \bibinfo{pages}{3475} (\bibinfo{year}{2001}).

\bibitem[{\citenamefont{{Spergel {\em et al.}}}(2003)}]{wmap:para}
\bibinfo{author}{\bibfnamefont{D.~N.} \bibnamefont{{Spergel {\em et al.}}}},
  \bibinfo{journal}{Astroph.\ J.\ Suppl.\ Ser.} \textbf{\bibinfo{volume}{148}},
  \bibinfo{pages}{175} (\bibinfo{year}{2003}).

\bibitem[{\citenamefont{{Copeland {\em et al.}}}(1994)}]{hybridinf:copeland}
\bibinfo{author}{\bibfnamefont{E.~J.} \bibnamefont{{Copeland {\em et al.}}}},
  \bibinfo{journal}{Phys.\ Rev.\ D} \textbf{\bibinfo{volume}{49}},
  \bibinfo{pages}{6410} (\bibinfo{year}{1994}).

\bibitem[{\citenamefont{Sarangi and Tye}(2002)}]{cosmicstrings_inf:sarangi}
\bibinfo{author}{\bibfnamefont{S.}~\bibnamefont{Sarangi}} \bibnamefont{and}
  \bibinfo{author}{\bibfnamefont{S.-H.~H.} \bibnamefont{Tye}},
  \bibinfo{journal}{Phys.\ Lett.\ B} \textbf{\bibinfo{volume}{536}},
  \bibinfo{pages}{185} (\bibinfo{year}{2002}).

\bibitem[{\citenamefont{Dvali and Vilenkin}(2004)}]{cosmicstrings_inf:dvali}
\bibinfo{author}{\bibfnamefont{G.}~\bibnamefont{Dvali}} \bibnamefont{and}
  \bibinfo{author}{\bibfnamefont{A.}~\bibnamefont{Vilenkin}},
  \bibinfo{journal}{JCAP} \textbf{\bibinfo{volume}{403}}, \bibinfo{pages}{10}
  (\bibinfo{year}{2004}).

\bibitem[{\citenamefont{Jeannerot et~al.}(2003)\citenamefont{Jeannerot, Rocher,
  and Sakellariadou}}]{cosmicstrings_inf:jeannerot}
\bibinfo{author}{\bibfnamefont{R.}~\bibnamefont{Jeannerot}},
  \bibinfo{author}{\bibfnamefont{J.}~\bibnamefont{Rocher}}, \bibnamefont{and}
  \bibinfo{author}{\bibfnamefont{M.}~\bibnamefont{Sakellariadou}},
  \bibinfo{journal}{Phys.\ Rev.\ D} \textbf{\bibinfo{volume}{68}},
  \bibinfo{pages}{103514} (\bibinfo{year}{2003}).

\bibitem[{\citenamefont{Rocher and
  Sakellariadou}(2004)}]{cosmicstrings_susygut}
\bibinfo{author}{\bibfnamefont{J.}~\bibnamefont{Rocher}} \bibnamefont{and}
  \bibinfo{author}{\bibfnamefont{M.}~\bibnamefont{Sakellariadou}},
  \bibinfo{journal}{hep-ph/0406120; hep-ph/0412143 (accepted by Phys.\ Rev.\
  Lett.)}  (\bibinfo{year}{2004}).

\bibitem[{\citenamefont{{Pogosian {\em et
  al.}}}(2003)}]{cosmicstrings:pogosian}
\bibinfo{author}{\bibfnamefont{L.}~\bibnamefont{{Pogosian {\em et al.}}}},
  \bibinfo{journal}{Phys.\ Rev.\ D} \textbf{\bibinfo{volume}{68}},
  \bibinfo{pages}{23506} (\bibinfo{year}{2003}).

\bibitem[{\citenamefont{Pogosian et~al.}(2004)\citenamefont{Pogosian, Wyman,
  and Wasserman}}]{cosmicstrings_wmap_sdss:pogosian}
\bibinfo{author}{\bibfnamefont{L.}~\bibnamefont{Pogosian}},
  \bibinfo{author}{\bibfnamefont{M.}~\bibnamefont{Wyman}}, \bibnamefont{and}
  \bibinfo{author}{\bibfnamefont{I.}~\bibnamefont{Wasserman}},
  \bibinfo{journal}{JCAP} \textbf{\bibinfo{volume}{9}}, \bibinfo{pages}{8}
  (\bibinfo{year}{2004}).

\bibitem[{\citenamefont{Bouchet et~al.}(1988)\citenamefont{Bouchet, Bennett,
  and Stebbins}}]{cosmicstrings:bouchet_nature}
\bibinfo{author}{\bibfnamefont{F.~R.} \bibnamefont{Bouchet}},
  \bibinfo{author}{\bibfnamefont{D.~P.} \bibnamefont{Bennett}},
  \bibnamefont{and} \bibinfo{author}{\bibfnamefont{A.}~\bibnamefont{Stebbins}},
  \bibinfo{journal}{Nature} \textbf{\bibinfo{volume}{335}},
  \bibinfo{pages}{410} (\bibinfo{year}{1988}).

\bibitem[{\citenamefont{Wu et~al.}(2002)\citenamefont{Wu, Avelino, Shellard,
  and Allen}}]{cosmicstrings:wu}
\bibinfo{author}{\bibfnamefont{J.~H.~P.} \bibnamefont{Wu}},
  \bibinfo{author}{\bibfnamefont{P.~P.} \bibnamefont{Avelino}},
  \bibinfo{author}{\bibfnamefont{E.~P.~S.} \bibnamefont{Shellard}},
  \bibnamefont{and} \bibinfo{author}{\bibfnamefont{B.}~\bibnamefont{Allen}},
  \bibinfo{journal}{Int.\ J.\ Mod.\ Phys.\ D} \textbf{\bibinfo{volume}{11}},
  \bibinfo{pages}{61} (\bibinfo{year}{2002}).

\bibitem[{\citenamefont{Wu}(1999)}]{phdthesis}
\bibinfo{author}{\bibfnamefont{J.~H.~P.} \bibnamefont{Wu}}, Ph.D. thesis,
  \bibinfo{school}{University of Cambridge, available at
  http://jhpw.phys.ntu.edu.tw} (\bibinfo{year}{1999}).

\bibitem[{\citenamefont{Landriau and Shellard}(2004)}]{cosmicstrings:cmb}
\bibinfo{author}{\bibfnamefont{M.}~\bibnamefont{Landriau}} \bibnamefont{and}
  \bibinfo{author}{\bibfnamefont{E.~P.~S.} \bibnamefont{Shellard}},
  \bibinfo{journal}{Phys.\ Rev.\ D} \textbf{\bibinfo{volume}{69}},
  \bibinfo{pages}{023003} (\bibinfo{year}{2004}).

\bibitem[{\citenamefont{Wu}(2004)}]{cosmicstrings_amiba}
\bibinfo{author}{\bibfnamefont{J.~H.~P.} \bibnamefont{Wu}},
  \bibinfo{journal}{Mod.\ Phys.\ Lett.\ A} \textbf{\bibinfo{volume}{19}},
  \bibinfo{pages}{1019} (\bibinfo{year}{2004}).

\bibitem[{\citenamefont{Seljak and Zaldarriaga}(1996)}]{cmbfast}
\bibinfo{author}{\bibfnamefont{U.}~\bibnamefont{Seljak}} \bibnamefont{and}
  \bibinfo{author}{\bibfnamefont{M.}~\bibnamefont{Zaldarriaga}},
  \bibinfo{journal}{Astroph.\ J.} \textbf{\bibinfo{volume}{469}},
  \bibinfo{pages}{437} (\bibinfo{year}{1996}).

\bibitem[{\citenamefont{{Albrecht {\em et al.}}}(1996)}]{defects_incoh}
\bibinfo{author}{\bibfnamefont{A.}~\bibnamefont{{Albrecht {\em et al.}}}},
  \bibinfo{journal}{Phys.\ Rev.\ Lett.} \textbf{\bibinfo{volume}{76}},
  \bibinfo{pages}{1413} (\bibinfo{year}{1996}).

\bibitem[{\citenamefont{Hu and Sugiyama}(1996)}]{cosperturb:hu}
\bibinfo{author}{\bibfnamefont{W.}~\bibnamefont{Hu}} \bibnamefont{and}
  \bibinfo{author}{\bibfnamefont{N.}~\bibnamefont{Sugiyama}},
  \bibinfo{journal}{Astroph.\ J.} \textbf{\bibinfo{volume}{471}},
  \bibinfo{pages}{542} (\bibinfo{year}{1996}).

\bibitem[{\citenamefont{{Hinshaw {\em et al.}}}(2003)}]{wmap:pow}
\bibinfo{author}{\bibfnamefont{G.}~\bibnamefont{{Hinshaw {\em et al.}}}},
  \bibinfo{journal}{Astroph.\ J.\ Suppl.\ Ser.} \textbf{\bibinfo{volume}{148}},
  \bibinfo{pages}{135} (\bibinfo{year}{2003}).

\bibitem[{\citenamefont{Bevis et~al.}(2004)\citenamefont{Bevis, Hindmarsh, and
  Kunz}}]{textures:wmap}
\bibinfo{author}{\bibfnamefont{N.}~\bibnamefont{Bevis}},
  \bibinfo{author}{\bibfnamefont{M.}~\bibnamefont{Hindmarsh}},
  \bibnamefont{and} \bibinfo{author}{\bibfnamefont{M.}~\bibnamefont{Kunz}},
  \bibinfo{journal}{Phys.\ Rev.\ D} \textbf{\bibinfo{volume}{70}},
  \bibinfo{pages}{043508} (\bibinfo{year}{2004}).

\bibitem[{\citenamefont{{Sazhin {\em et
  al.}}}(2003)}]{cosmicstrings_lens:sazhin}
\bibinfo{author}{\bibfnamefont{M.}~\bibnamefont{{Sazhin {\em et al.}}}},
  \bibinfo{journal}{Mon.\ Not.\ R.\ Astron.\ Soc.}
  \textbf{\bibinfo{volume}{343}}, \bibinfo{pages}{353} (\bibinfo{year}{2003}).

\bibitem[{\citenamefont{{Schild {\em et
  al.}}}(2004)}]{cosmicstrings_lens:schild}
\bibinfo{author}{\bibfnamefont{R.}~\bibnamefont{{Schild {\em et al.}}}},
  \bibinfo{journal}{Astron.\ Astrophys.} \textbf{\bibinfo{volume}{422}},
  \bibinfo{pages}{477} (\bibinfo{year}{2004}).

\end{thebibliography}

\end{document}